\documentclass[final,english]{bullsrsl}[2022/06/15]

\usepackage[latin1]{inputenc}
\usepackage[T1]{fontenc}
\usepackage{amsmath}

\usepackage{natbib} 
\usepackage{amsmath,amsfonts,amssymb}
\hypersetup{colorlinks=true, allcolors=blue}
\usepackage{booktabs}
\usepackage{siunitx}
\usepackage{graphicx}%
\usepackage{multirow}%
\usepackage{mathrsfs}%
\usepackage[title]{appendix}%
\usepackage{xcolor}%
\usepackage{textcomp}%
\usepackage{manyfoot}%
\usepackage{booktabs}%
\usepackage{algorithm}%
\usepackage{algorithmicx}%
\usepackage{algpseudocode}%
\usepackage{listings}%
\usepackage{xspace}
\usepackage{subcaption}
\usepackage{tikz}
\usepackage{siunitx}
\usepackage{caption}

\usepackage{array}

\begin{document}
\title{Design and Development of a Lab Prototype of a Fiber-Based Integral Field Spectrograph
}

\author[affil={1}, corresponding]{Nitish}{Singh}
\author[affil={1}]{S.}{Sriram}
\author[affil={2}]{Jurgen}{Schmoll}
\author[affil={1}]{Bharat Kumar}{Yerra}
\affiliation[1]{Indian Institute of Astrophysics, Koramangala, Bengaluru, 560034, INDIA}
\affiliation[2]{Centre for Advanced Instrumentation, Durham University, United Kingdom}
\correspondance{nitiphy@gmail.com, nitish.singh@iiap.res.in}
\maketitle

\begin{abstract}
Integral Field Spectroscopy provides simultaneous spatial and spectral information, making it a powerful technique for studying both point like and extended astronomical sources. As part of our effort to develop a compact Integral Field Spectrograph (IFS) for optical astronomy, we have designed and realized a fiber based prototype with a resolving power of $\sim$800 at the H-$\alpha$ wavelength. The front end optics, including the fore optics and lenslet based integral field unit (IFU), have been designed and optimized. In the present work, we focus on the design and laboratory verification of a fiber only IFU module consisting of 37 fibers arranged in a precise hexagonal geometry to match the lenslet pitch. The back end of the system forms a linear fiber slit that feeds a laboratory built spectrograph. The optical design was optimized using ZEMAX, and the system was experimentally tested using a Neon emission lamp and solar light. The measured spectra clearly show prominent features including the H-$\alpha$ lines, with dispersion and resolution consistent with theoretical predictions. Although lenslets are not yet integrated, this work establishes a validated methodology for accurate fiber alignment for IFS. This results form a crucial step toward the full implementation of a lenslet fiber based IFS for the 2.34 meter Vainu Bappu Telescope.

\end{abstract}


\section{Introduction}

Integral Field Spectrograph (IFS) is a powerful observational instrument that enables the simultaneous acquisition of spatially resolved spectra across an extended field of view. By capturing both spatial and spectral information in a single exposure, IFS allows detailed studies of the physical, chemical, and kinematic properties of astronomical sources. In an IFS, the Integral Field Unit (IFU) serves as the front end interface that divides the telescope focal plane into multiple spatial elements and feeds the collected light into a spectrograph \citep{2002PASP..114..866R}. Fiber-based IFUs are particularly attractive due to their mechanical flexibility, ease of alignment, and compatibility with compact spectrograph designs. As part of an ongoing effort to develop a lenslet fiber based IFS for the prime focus of the 2.34 m Vainu Bappu Telescope (VBT) \citep{2024SPIE13100E..4OS}, we are developing and validating the key subsystems of the instrument. In the complete system, a microlens array will sample the prime focus focal plane and couple light into optical fibers. In the present work, however, we focus exclusively on the design, fabrication, and laboratory validation of the fiber only IFU module, prior to integrating the lenslet array. The IFU consists of 37 optical fibers arranged in a precise hexagonal geometry, with the fiber pitch matched to the designed microlens array pitch. This staged approach allows independent verification of fiber positioning accuracy, slit reformatting, and spectrograph performance. Each fiber collects light from a distinct spatial location and reformat the 2D input field into a 1D linear slit at the spectrograph entrance. The spectrograph disperses the light along the wavelength (dispersion) axis while preserving spatial information along the orthogonal axis, producing a three dimensional data cube. The fiber only IFU and spectrograph were designed and optimized using ZEMAX and experimentally tested in the laboratory using neon emission lamps and solar light.

\section{Optical Design of the Lenslet and Fore Optics}

The prime focus of the VBT delivers an f/3.25 beam, which is corrected using a wide field corrector (WFC) developed in our laboratory to provide an effective f/3.42 beam \citep{2025ExA....60....3S}. Based on this optical configuration, the fore optics and lenslet array were designed to meet the requirements of the IFS. The detailed optical design and performance analysis of the fore optics system have been presented earlier in \cite{2024SPIE13100E..4OS}. Although the original design was optimized for the OMR Spectrograph (OMRS), the same front end optical configuration has been adapted for the present lenslet fiber based IFS. The lenslet array provides a nearly telecentric pupil at the pupil plane with an effective f/ratio of 3.67. The resulting spot size at the focal plane is smaller than $90~\mu$m. To efficiently collect the incident light, a fiber with a core diameter of $100~\mu$m and a numerical aperture (NA) of 0.22 is employed, ensuring minimal coupling losses.

\begin{figure}[htbp]

    \centering
    \includegraphics[width=0.7\linewidth]{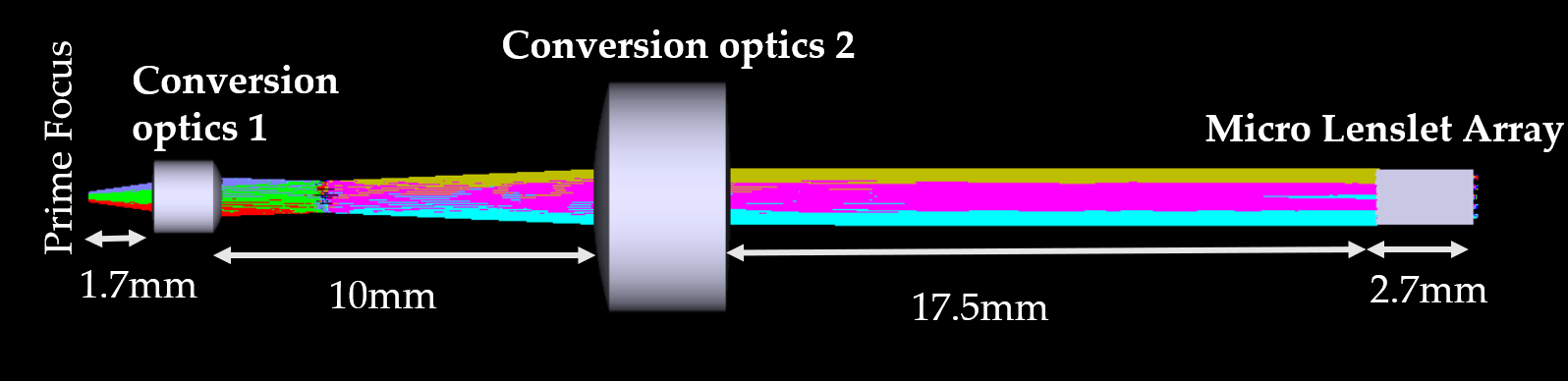}
    \caption{OMRS optical design for Prime mode with lenslet}
    \label{fig:design_lenslet}

\end{figure}

Figure~\ref{fig:design_lenslet} shows the OMRS prime focus optical design with the lenslet array. In this work, we focus on the fiber arrangement at the lenslet array pupil plane. The pitch of the fabricated lenslet array was experimentally verified using a photometric setup. The measured pitch was $507 \pm 2~\mu$m, and the focal spots spanned approximately 5 pixels on the CMOS detector, with a variation of $\pm 1$ pixel, confirming good agreement with the design.
\begin{figure}[htbp]
    \centering

    \begin{subfigure}[b]{0.55\linewidth}
        \centering
        \includegraphics[width=\textwidth]{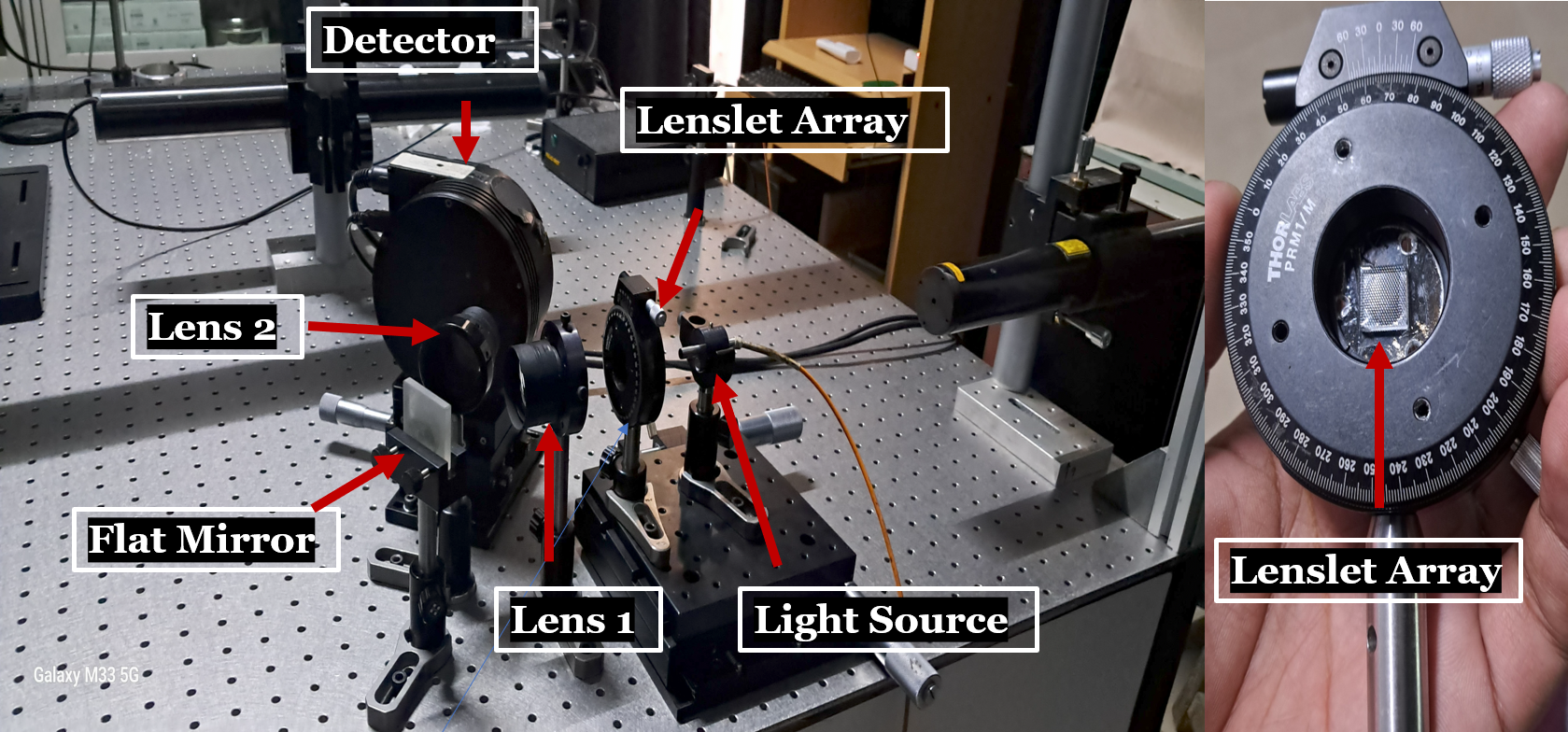}
        \caption{}
        \label{fig:photometry}
    \end{subfigure}
    \hspace{0.02\linewidth} 
    \begin{subfigure}[b]{0.25\linewidth}
        \centering
        \includegraphics[width=\textwidth]{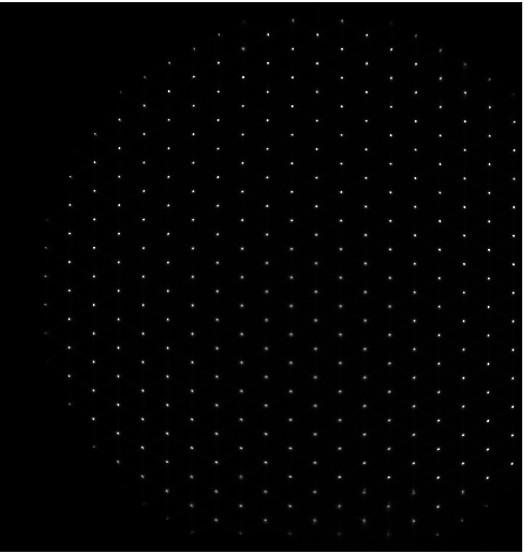}
        \caption{}
        \label{fig:lensletimage}
    \end{subfigure}

    \caption{Lenslet imaging verification: (a) laboratory imaging setup used to characterize the lenslet, (b) pupil image recorded at the detector.}
    \label{fig:lenslet_all}
\end{figure}

\section{Fiber Arrangement Methodology}

\begin{figure}[htbp]
    \centering
    \begin{subfigure}[b]{0.48\linewidth}
        \centering
        \includegraphics[width=\linewidth]{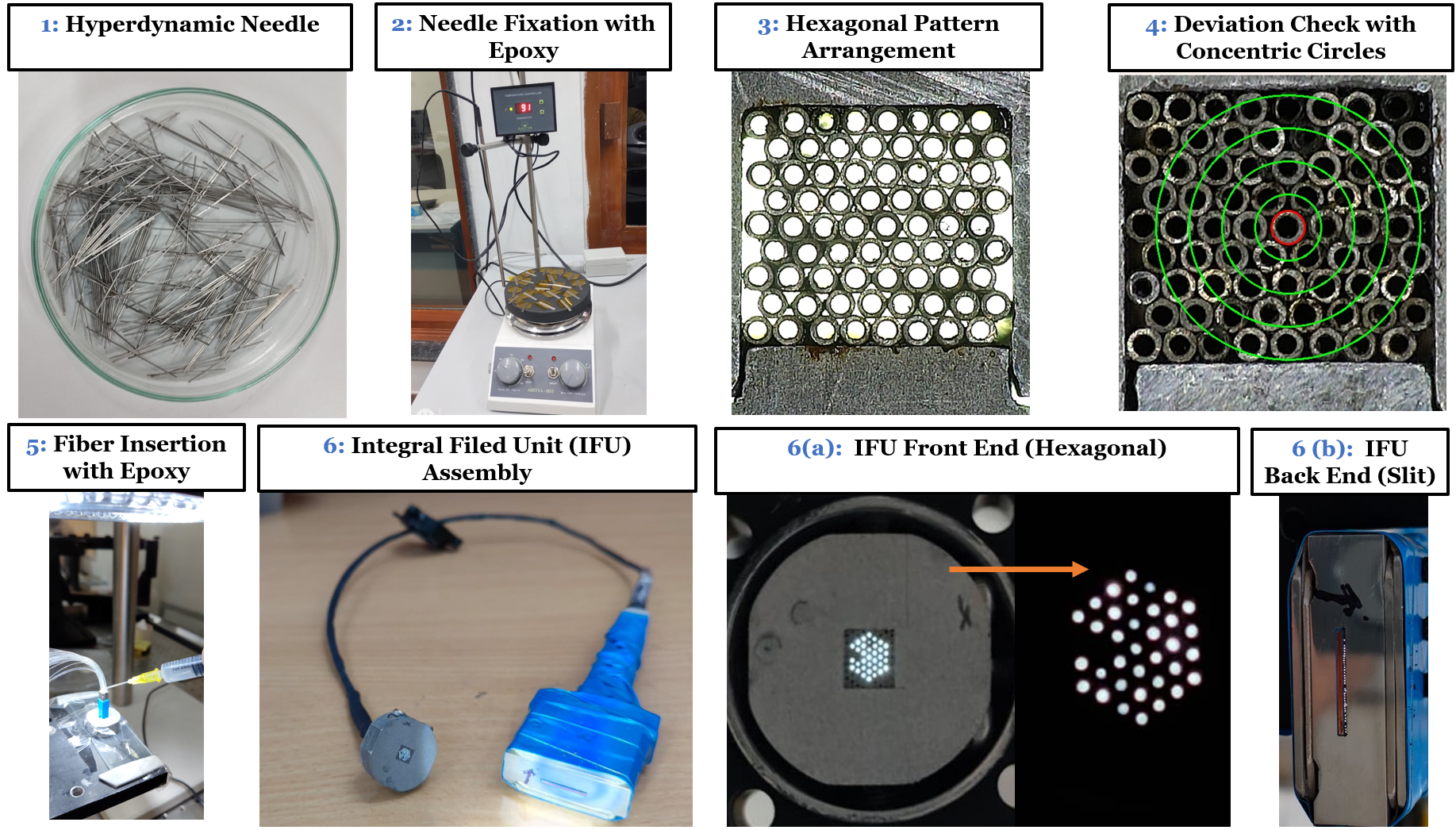}
        \caption{}
        \label{fig:ifuassembly_febrication}
    \end{subfigure}
    \hfill
    \begin{subfigure}[b]{0.50\linewidth}
        \centering
        \includegraphics[width=\linewidth]{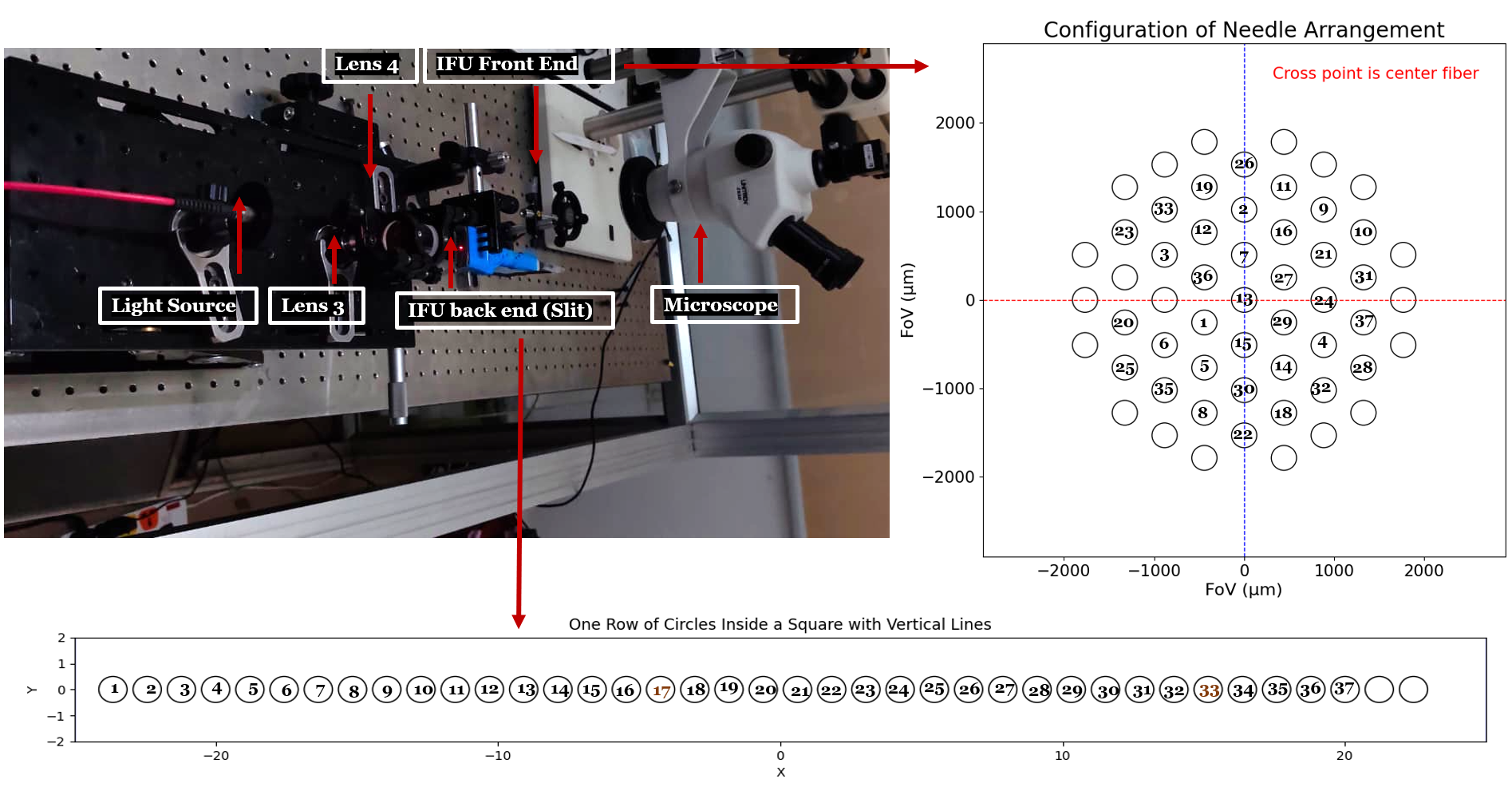}
        \caption{}
        \label{fig:ifu_alignment}
    \end{subfigure}
    \caption{Fabrication and alignment verification of the IFU, (a) IFU fabrication methodology. (b) IFU fiber scanning and numbering }
    \label{fig:ifu_fabrication_alignment}
\end{figure}

After verifying the pitch of each lenslet pupil, we proceeded to arrange the optical fibers in a hexagonal geometry matching the lenslet pitch. This step required positioning all 100 $\mu m$ core fibers with a target alignment accuracy better than 5 $\mu m$ relative to the lenslet pitch. Fiber alignment is carried out using a stainless-steel hypodermic needle based method developed by \cite{2004SPIE.5492..624S}. Approximately 400 needles are individually measured in the laboratory using a screw gauge with 1 $\mu m$ precision, and those with outer diameters in the range 506-515 $\mu m$ are selected for assembly. The selected needles have a nominal outer diameter of 510 $\mu m$ and an inner diameter of 250 $\mu m$, making them suitable for fibers with a 240 $\mu m$ polymer jacket. This selection ensures proper packing within the nominal 507 $\mu m$ pitch arrangement, closely matching the lenslet pitch. The needles were arranged in a hexagonal configuration and bonded using Epotech-2 epoxy for structural stability and quick curing. The alignment accuracy of the array was verified using concentric circle overlays to ensure the deviation remained within tolerance. Subsequently, optical fibers were inserted into the aligned needles, and their back ends were secured in a V-groove shaped slit. The fibers were also glued in place using Epotech-2 epoxy to ensure mechanical stability. Finally, the overall fiber positioning accuracy achieved in the assembled IFU is $<50~\mu m$ relative to the lenslet pitch. The resulting IFU assembly featured a front end with a hexagonal geometry matched to the lenslet array and back end in slit form, as illustrated in Figure~\ref{fig:ifuassembly_febrication}.

After inserting all fibers into the hypodermic needles and securing them in the V-groove slit, we verified fiber alignment and continuity.A 200~$\mu m$ test fiber was illuminated at the focal plane of lens 3 (25~mm diameter, 100~mm focal length) to produce a collimated beam, which was then reduced to 100~$\mu m$ using a demagnifier lens (lens 4; 25~mm diameter, 50~mm focal length) and focused onto the IFU slit end. The slit, mounted on an X-Y-Z stage with 10~$\mu m$ accuracy, was scanned while observing the front end hexagonal fiber pattern under a microscope. Each fiber was imaged individually to map the slit fibers to their corresponding lenslets. Out of 37 fibers, 35 were successfully illuminated, while fibers 17 and 33 showed no signal, likely due to breakage during fabrication, as shown in Figure~\ref{fig:ifu_alignment}.

\section{Imaging Setup for IFU Testing}

\begin{figure}[htbp]

    \centering
    \includegraphics[width=0.8\linewidth]{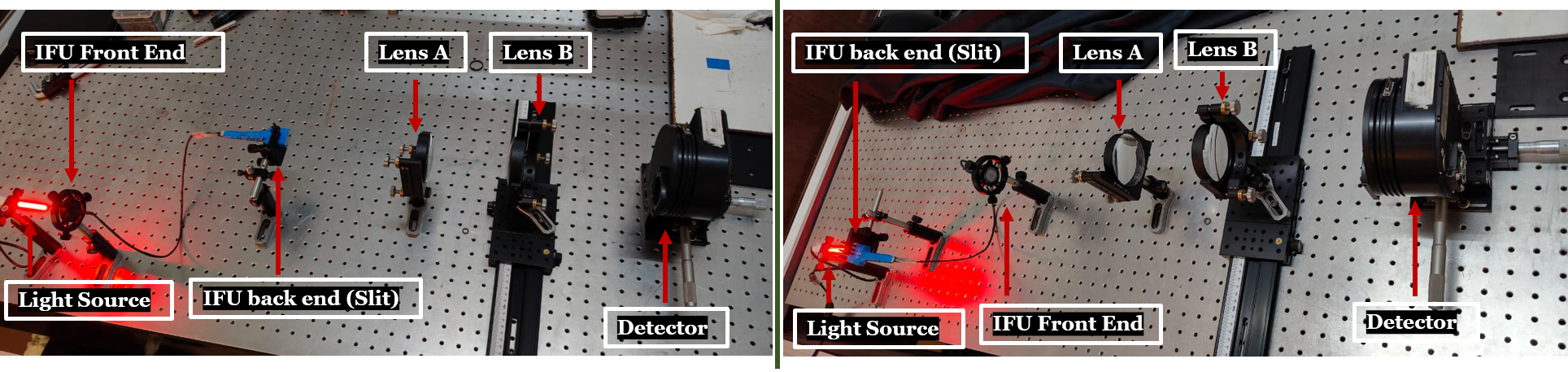}
    \caption{ IFU imaging laboratory setup. Left panel: schematic of the optical configuration used for testing the IFU front and back ends. Right panel: experimental laboratory setup showing the alignment of the IFU, lenses, light source, and detector.}
    \label{fig:ifu_imaging_setup}

\end{figure}

To test the front and back ends of the IFU, a dedicated imaging setup was constructed. The setup employed two lenses, Lens A and Lens B, arranged in a one to one imaging configuration. The collimator and camera lenses (Lens A and Lens B) were achromatic lenses with 75 mm diameter and 200 mm focal length, suitable for VIS-NIR operation. A neon lamp was used as the light source. The IFU front end was positioned at the focal plane of Lens A, while the back-end slit of the IFU was illuminated by the light source. Lens A collimated the beam from the IFU, which was then focused by Lens B onto the CMOS detector having an active area of ($0.5 \times 0.5$) pixel with a pixel size of 20 $\mu m$/pixel. The same optical setup was also used to image the opposite side of the IFU by interchanging the front-end and back-end positions at the focal plane. The setup is illustrated in Figure~\ref{fig:ifu_imaging_setup}.

\begin{figure}[htbp]

    \centering
    \includegraphics[width=0.8\linewidth]{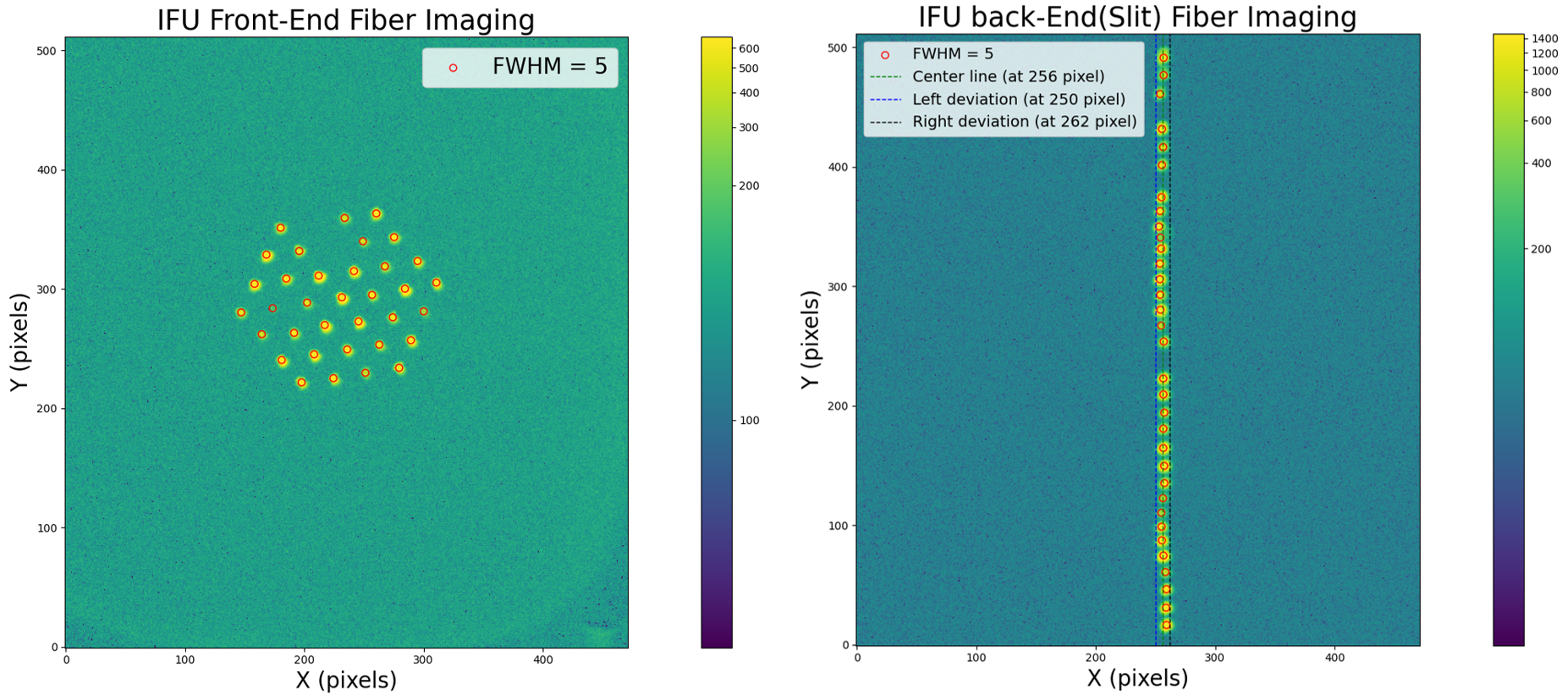}
    \caption{ Result of IFU Imaging Lab Setup}
    \label{fig:ifu_lab_results}
\end{figure}

After setting up the imaging system, laboratory tests were performed on the IFU. Both the front and back ends of the IFU were imaged using the same optical setup described earlier (Figure~\ref{fig:ifu_imaging_setup}), and a total of 35 fibers were successfully illuminated. The setup allowed illumination of both sides of the IFU face using the same CMOS detector. Each fiber produced a spot covering approximately five pixels on the detector, corresponding to a diameter of 100 $\mu m$, which matches the fiber core size. At the slit end, the maximum deviation from the nominal center position was measured to be $\pm$6 pixels from the first to the last fiber. The results of this test are shown in Figure~\ref{fig:ifu_lab_results}. The setup allowed us to illuminate the fiber bundle and capture images from both the hexagonal input face and the linear slit output face. The recorded images confirmed that the output spots from each fiber were clearly resolved, with a typical FWHM of ~5 pixels measured using ApertureStats, corresponding to a spot diameter of ~100~$\mu m$ at an image scale of 20~$\mu m$/pixel, consistent with the fiber core size. However, two fibers were found to be damaged during handling, resulting in detectable light from 35 out of 37 fibers, as shown in Figure \ref{fig:ifu_lab_results}.

\setlength{\intextsep}{0pt}
\setlength{\intextsep}{0pt}

\section{IFS Design, Experimental Setup, and Spectra}

\begin{figure}[htbp]
    \centering

    \begin{subfigure}[b]{0.46\linewidth}
        \centering
        \includegraphics[width=\textwidth]{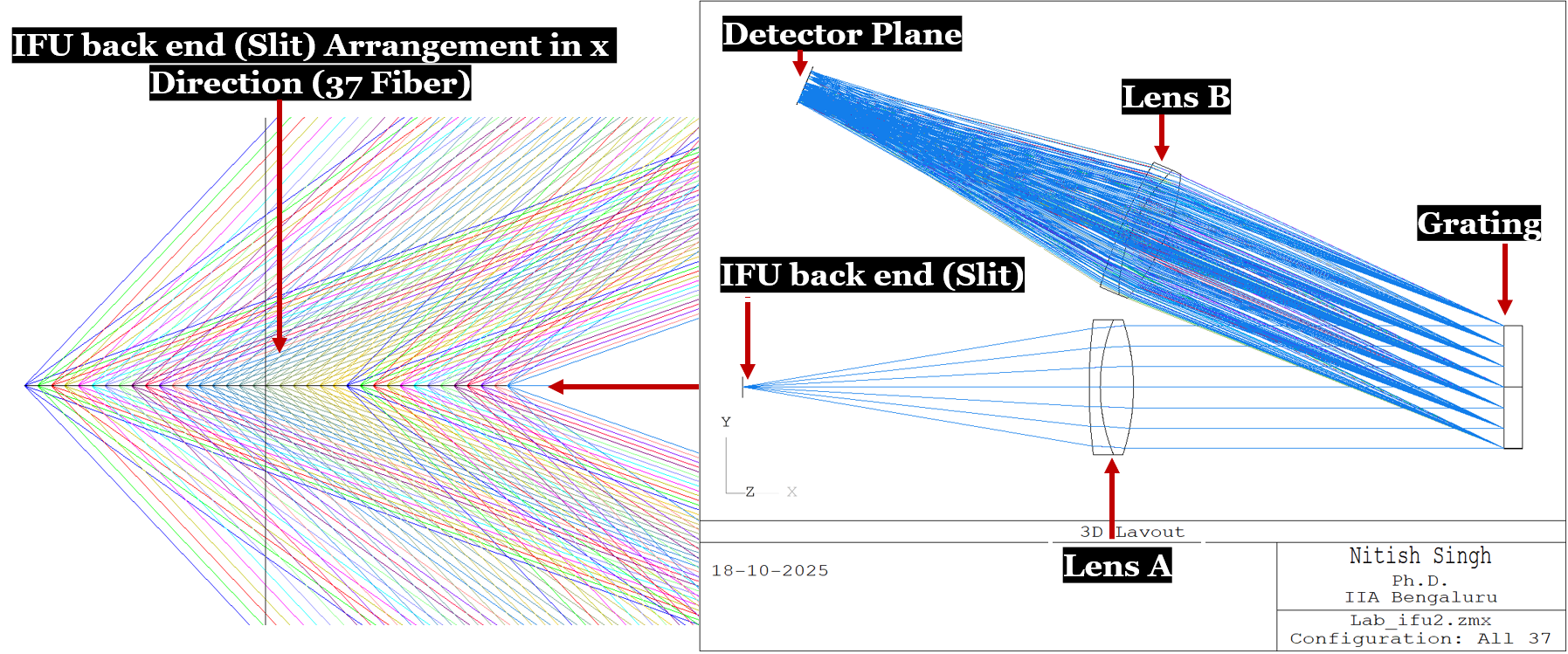}
        \caption{}
        \label{fig:fiber_slit_simulation}
    \end{subfigure}
    \hspace{0.02\linewidth} 
    \begin{subfigure}[b]{0.44\linewidth}
        \centering
        \includegraphics[width=\textwidth]{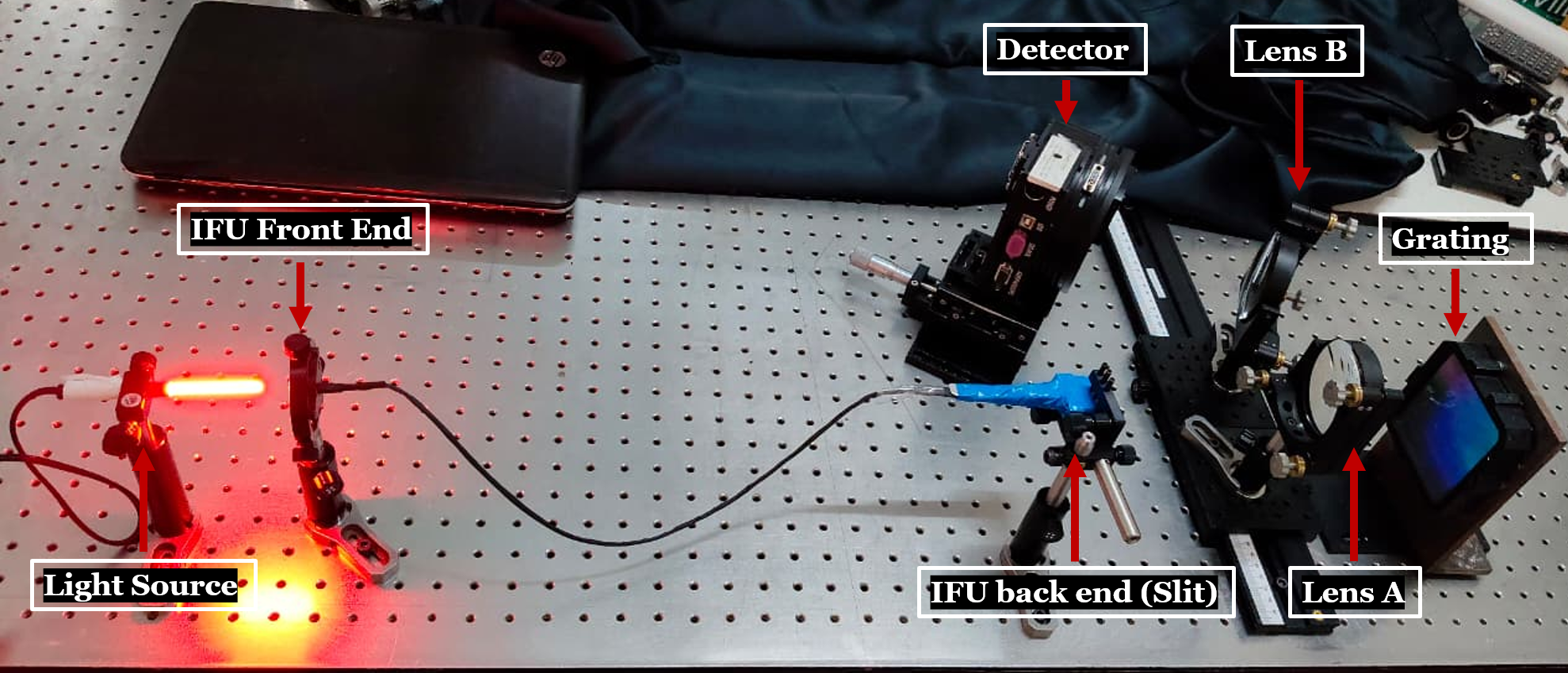}
        \caption{}
        \label{fig:lab_spectrograph}
    \end{subfigure}

    \caption{(a) Integral Field Spectrograph optical layout in ZEMAX, (b) Laboratory setup of the Integral Field Spectrograph.}
    \label{fig:IFU_spectrograph_all}
\end{figure}

\begin{figure}[htbp]
  \centering

  \begin{subfigure}[b]{0.36\linewidth}
    \centering
    \includegraphics[width=\textwidth]{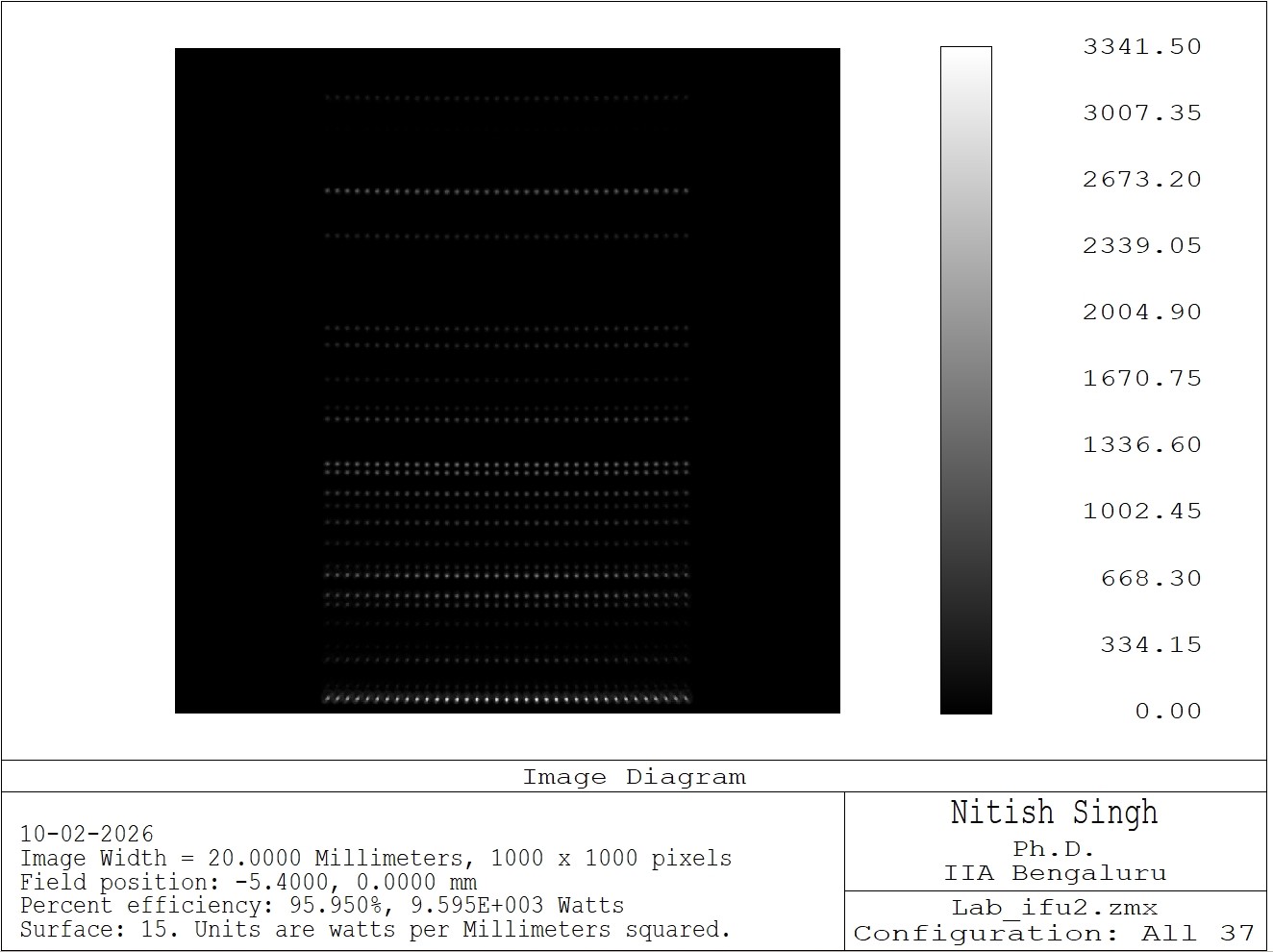}
    \caption{}
    \label{fig:2D_Spectra_Neon_ZEMAX_direct}
  \end{subfigure}
    \hspace{0.02\linewidth} 
  \begin{subfigure}[b]{0.29\linewidth}
    \centering
    \includegraphics[width=\textwidth]{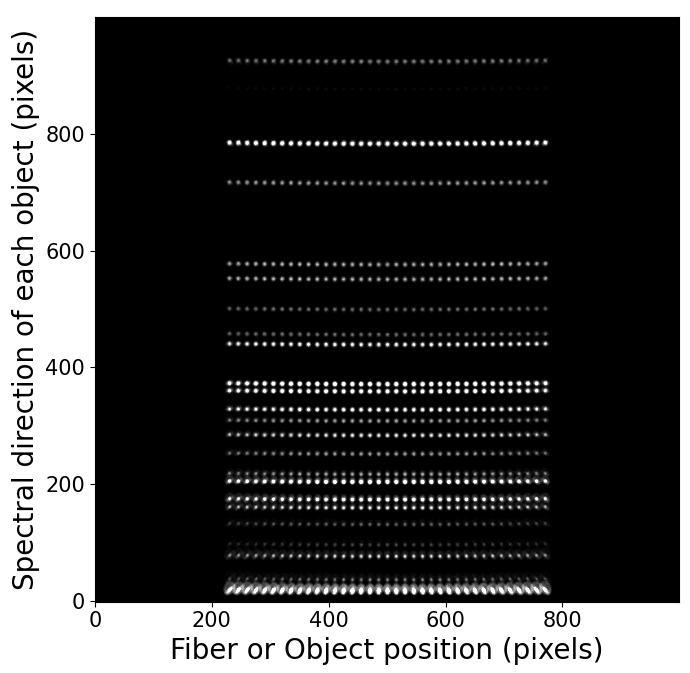}
    \caption{}
    \label{fig:Neon_ZEMAX_python}
  \end{subfigure}
    \hspace{0.02\linewidth} 
  \begin{subfigure}[b]{0.28\linewidth}
    \centering
    \includegraphics[width=\textwidth]{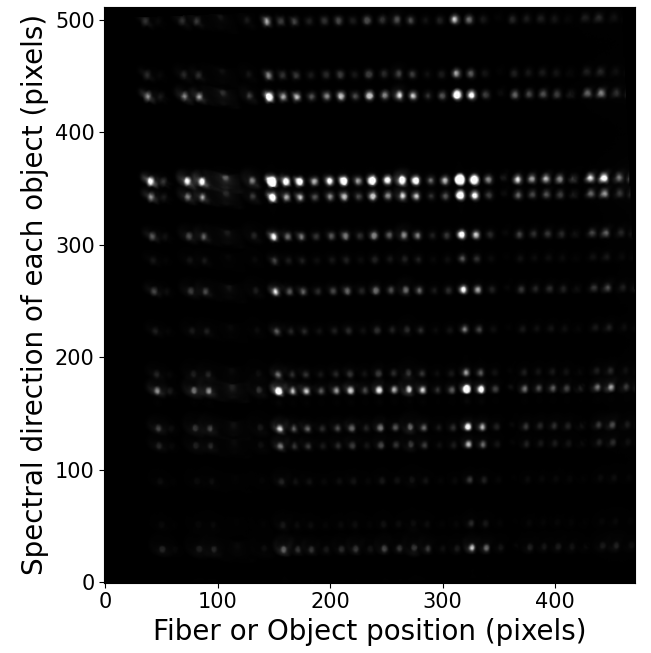}
    \caption{}
    \label{fig:Neon_Lab}
  \end{subfigure}

  \caption{2D spectra at the IFU slit end for Neon lamp emission. 
  (a) Geometrical image simulation from ZEMAX, 
  (b) ZEMAX simulation reconstructed in Python, 
  (c) Measured in the laboratory.}
  
  \label{fig:2D_Spectra_all}
\end{figure}

The spectrograph was designed and optimized in ZEMAX over the wavelength range 0.4-0.7 $\mu m$. In the simulations, neon lamp emission lines between 5852.49 and 7400~\AA were used, with weights assigned according to their relative intensities. All 37 fibers were arranged in a linear slit configuration with a core to core spacing of 300~$\mu$m to replicate the actual fiber bundle used in the instrument (Figure~\ref{fig:fiber_slit_simulation}). Following the optical design, the spectrograph was assembled in the laboratory using Collimator Lens (Lens~A) and Camera lens (Lens~B), both achromatic doublets with a focal length of 200~mm and a clear aperture of 75~mm, along with a 600~lines/mm reflection grating. The system employed optical fibers with a core diameter of 100~$\mu$m and a 512$\times$512 pixel detector with a pixel size of 20~$\mu$m to record the spectra (Figure~\ref{fig:lab_spectrograph}). Using this setup, two dimensional spectra of a neon lamp were first simulated in ZEMAX to study the spatial and spectral distribution at the IFU slit end. A geometrical image simulation was performed for the neon source (Figure~\ref{fig:2D_Spectra_Neon_ZEMAX_direct}), and the simulated spectra were reconstructed in Python (Figure~\ref{fig:Neon_ZEMAX_python}) for direct comparison with laboratory measurements. 

\begin{figure}[htbp]
  \centering
  \begin{subfigure}[b]{0.5\linewidth}
    \centering
    \includegraphics[width=\textwidth]{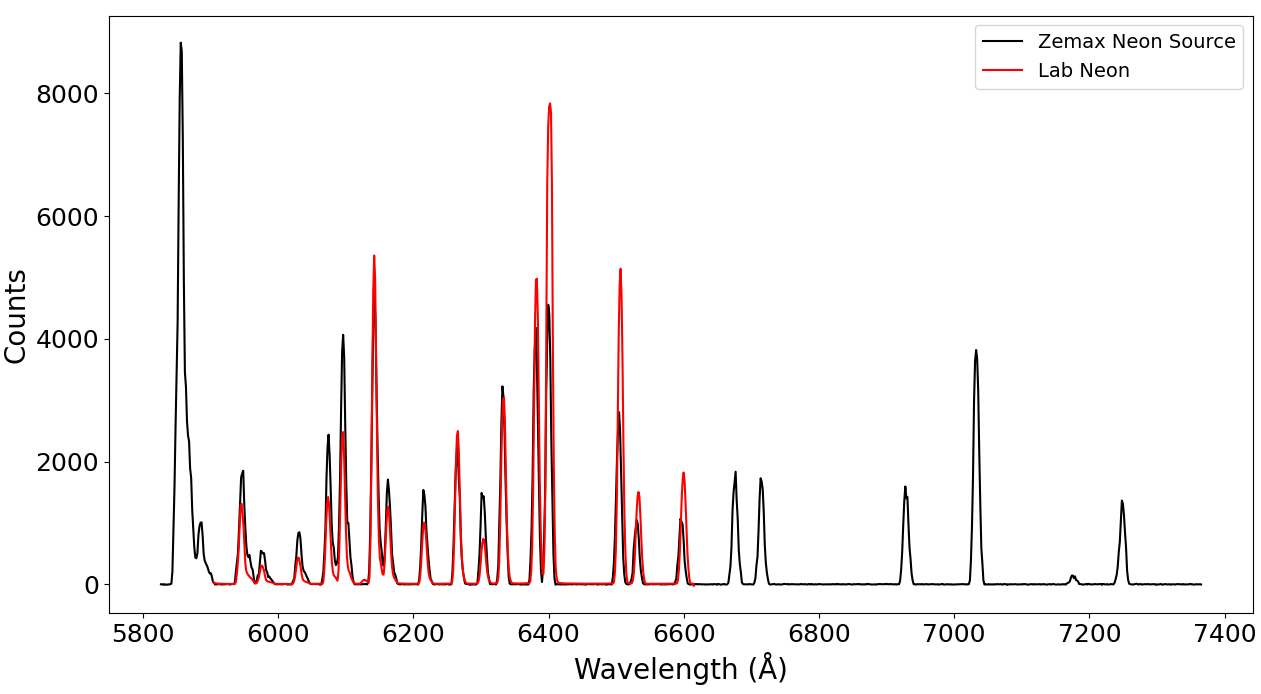}
    \caption{}
    \label{fig:Wavelegth_neon}
  \end{subfigure}
  \hspace{0.02\linewidth}
  \begin{subfigure}[b]{0.44\linewidth}
    \centering
    \includegraphics[width=\textwidth]{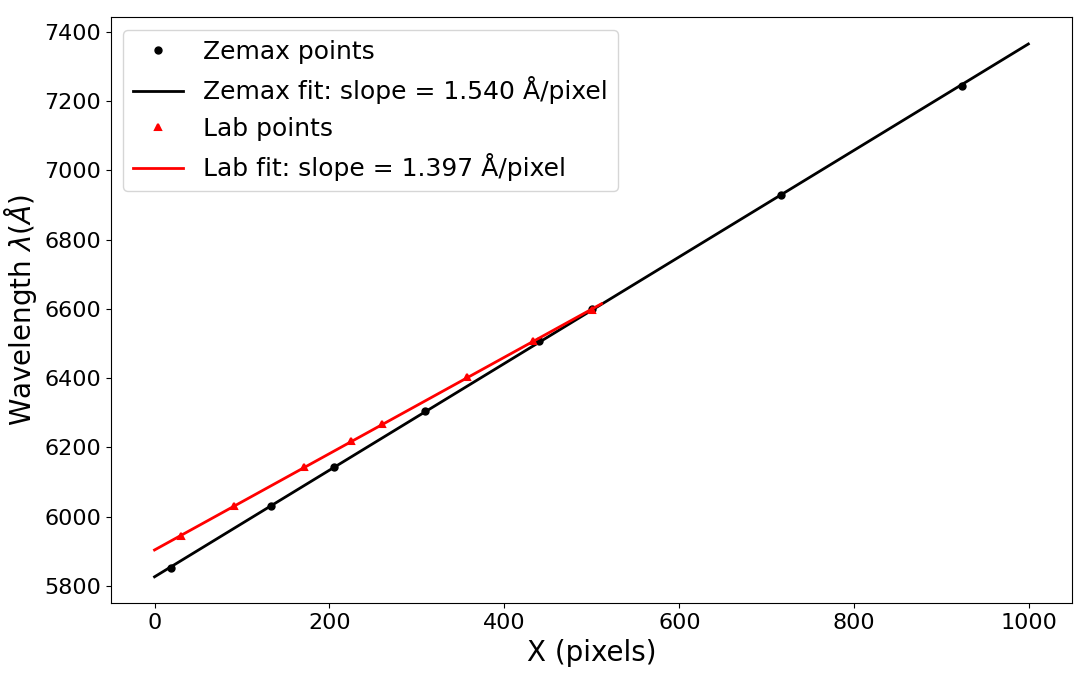}
    \caption{}
    \label{fig:dispersion_all}
  \end{subfigure}
\caption{(a) Wavelength Calibrated Neon lamp spectra from ZEMAX simulations and laboratory measurements for a single object. (b) Spectral dispersion (Linear dispersion versus wavelength) from ZEMAX and laboratory data for a single object.}
  \label{fig:Wavelegth_Space_lab_source}
\end{figure}

In the laboratory, the IFS recorded 2D neon spectra on the detector, clearly showing the corresponding spatial and spectral distribution of the emission lines (Figure~\ref{fig:Neon_Lab}). The 2D data were reduced to 1D spectra using a custom Python based pipeline employing standard scientific packages including Astropy, NumPy, SciPy, and Matplotlib. The extracted spectra were wavelength calibrated using identified neon emission lines, and a linear fit between wavelength and pixel position was used to derive the dispersion. For a representative single fiber, the measured dispersion is $1.54~\text{\AA/pixel}$ for the ZEMAX simulations and $1.39~\text{\AA/pixel}$ for the laboratory data (Figures~\ref{fig:Wavelegth_neon} and~\ref{fig:dispersion_all}).

\setlength{\intextsep}{0pt}
\setlength{\intextsep}{0pt}

\setlength{\intextsep}{0pt}
\setlength{\intextsep}{0pt}

\section{Spectral Resolution And Linear Dispersion of IFS }

The spectral performance of the IFS was estimated using three complementary methods: theoretical calculations, Zemax simulations, and laboratory measurements. Two widely separated Neon emission lines ($\lambda = 6402.25$~\AA, $\lambda_a = 6382.99$~\AA, $\delta \lambda = 19.26$~\AA) were used for these estimates (see Figure\ref{fig:Wavelegth_Space_lab_source}).

\subsection{Theoretical Estimation of Spectral Resolution and Linear Dispersion }

The theoretical spectral resolution of the IFS was estimated for a representative neon lamp emission line using the grating equation (shown in Equation~\ref{eq:grating_equation}). The calculation employs the same optical and detector parameters as those used in the laboratory setup, including the grating groove density($\sigma$), focal lengths of the collimator ($f_{\text{col}}$)  and camera ($f_{\text{cam}}$) , detector pixel size, and spectral sampling ($\Delta x$). This approach provides a sampling limited estimate of the achievable spectral resolution at the selected wavelength. The parameters used for the estimation are listed in Table~\ref{tab:theoretical_params}.  To define the relationship between the diffraction order ($m$), wavelength ($\lambda$), groove spacing ($d$), incident angle ($\alpha$), and diffraction angle ($\beta$), we use the Equations~\ref{eq:grating_derivation}.
{
\setlength{\abovedisplayskip}{0pt}
\setlength{\belowdisplayskip}{0pt}
\begin{subequations}
\label{eq:grating_derivation}
\begin{align}
    & m \lambda = d (\sin \alpha + \sin \beta) \label{eq:grating_equation} \\
    & \beta = \arcsin\left(\frac{m\lambda}{d} - \sin \alpha\right), \quad 
    \Delta\beta = \arctan\left(\frac{\Delta x}{f_{\text{cam}}}\right), \quad 
    \Delta\lambda = d \cos\beta \, \Delta\beta, \quad 
    R = \frac{\lambda}{\Delta\lambda} \label{eq:derived_params}
\end{align}
\end{subequations}
}

\setlength{\intextsep}{2pt}
\setlength{\intextsep}{2pt}
\begin{table}[htbp]
\centering
\setlength{\abovecaptionskip}{2pt} 
\setlength{\belowcaptionskip}{2pt}

\begin{minipage}[t]{0.48\textwidth}
    \centering
    \caption{Laboratory Spectrograph Parameters and Theoretical Resolution Estimation}
    \label{tab:theoretical_params}
    \renewcommand{\arraystretch}{0.9} 
    \begin{tabular}{l@{\hspace{0.5em}}l}
    \toprule
    \textbf{Parameter} & \textbf{Value} \\
    \midrule
    Wavelength ($\lambda$) & $6402.25~\text{\AA}$ \\
    Groove Density ($\sigma$) & $600~\text{lines/mm}$ \\
    Groove Spacing ($d$) & $1.667 \times 10^{-3}~\text{mm}$ \\
    $f_{\text{cam}}$ & $200~\text{mm}$ \\
    Pixel Size & $20~\mu\text{m}$ \\
    Core Dia & $100~\mu\text{m}$ \\
    Spectral Sampling ($\Delta x$) & $0.1~\text{mm}$ \\
    Diffraction Angle ($\beta$) & $22.5900^\circ$ \\
    Angular Sampling ($\Delta\beta$) & $4.996 \times 10^{-4}~\text{rad}$ \\
    Spectral Res. ($\Delta\lambda$) & $\approx 7.685~\text{\AA}$ \\
    Resolving Power ($R$) & $\approx 833$ \\
    \bottomrule
    \end{tabular}
\end{minipage}
\hfill
\begin{minipage}[t]{0.48\textwidth}
    \centering
    \caption{Verification of Spectral Dispersion and Resolving Power.}
    \label{tab:dispersion_verification}
    \renewcommand{\arraystretch}{0.9}
    \begin{tabular}{l@{\hspace{0.5em}}l}
    \toprule
    \textbf{Metric} & \textbf{Value} \\
    \midrule
    Wavelength Sep. ($\delta\lambda$) & $19.26~\text{\AA}$ \\
    Angular Sep. ($\Delta\beta_{sep}$) & $0.0717^\circ$ \\
    Linear Sep. ($\Delta X_{sep}$) & $250.28~\mu\text{m}$ \\
    $Dispersion_{theory}$ & $1.539~\text{\AA/px}$ \\
    FWHM (Sampling) & $5~\text{pixels}$ \\
    Theor. FWHM ($\Delta\lambda_{\text{theory}}$) & $7.695~\text{\AA}$ \\
    $R_{\text{theory}}$ & $\approx 832$ \\
    \bottomrule
    \end{tabular}
\end{minipage}
\end{table}

After estimating the spectral resolution theoretically (see Table~\ref{tab:theoretical_params}), the linear dispersion of the IFS were determined.  Two widely separated Neon emission lines ($\lambda = 6402.25$~\AA, $\lambda_a = 6382.99$~\AA, $\delta \lambda = 19.26$~\AA) were used for these estimation (as calculated in Table~\ref{tab:dispersion_verification}).
\[
\begin{aligned}
& \text{The linear separation between $\lambda$ and $\lambda_{a}$ on the detector is given by}  (\Delta X_{\text{theory}}) = f_{\text{camera}} \times \sin(\Delta \beta)
\quad
\\ &
\text{Theoretically Estimation of Linear Dispersion} (\text{Dispersion}_{\text{theory}}) = \frac{\lambda - \lambda_{a}}{\Delta X_{\text{pixel, theory}}}
\quad
\\ &
\text{Theoretically FWHM of the emission line} (\text{FWHM}_{\text{theory}}) = \frac{\text{Fiber Core Dia}}{\text{Pixel Size}} = \frac{100 \mu m}{20 \mu m} = \text{5}
\quad
\\&\text{Theortical Spectral Resolution} (\Delta\lambda_{\text{theory}}) = \text{FWHM}_{\text{theory}} \times \text{Dispersion}_{\text{theory}}
\end{aligned}
\]

\subsection{ZEMAX Simulation Estimation of Spectral Resolution and Linear Dispersion}

The spectrograph was designed and optimized in ZEMAX, as illustrated in Figure~\ref{fig:fiber_slit_simulation}. A total of 37 fibers (object fields) were configured at the slit plane to simulate the IFU output. For the dispersion analysis, the 2D simulated spectrum corresponding to a single fiber was extracted and wavelength calibrated, as shown in Figure~\ref{fig:Wavelegth_Space_lab_source}.
The ZEMAX simulation results, including detector positions for wavelengths $\lambda$ and $\lambda_a$ and the resulting spectral performance metrics, are derived as follows:

\noindent
\begin{minipage}{\textwidth}
\abovedisplayskip=0pt
\belowdisplayskip=0pt
\begin{gather*}
    \text{Detector pixel positions for $\lambda$ and $\lambda_a$ is: } X_{\text{ZEMAX}} = -2.550~\text{mm}, \quad X_{a,\text{ZEMAX}} = -2.799~\text{mm} \\
    \text{Centroid separation between $\lambda$ and $\lambda_a$} (\Delta X_{\text{ZEMAX}}) = |X_{a,\text{ZEMAX}} - X_{\text{ZEMAX}}| = 0.249~\text{mm} = 12.45~\text{pixels} \\
    \text{Linear dispersion for ZEMAX data} (\text{Dispersion}_{\text{ZEMAX}}) = \frac{\lambda - \lambda_a}{\Delta X_{\text{ZEMAX}}} = \frac{19.26~\text{\AA}}{12.45~\text{pixels}} = 1.546~\text{\AA/pixel} \\
    \text{FWHM (\text{determined from laboratory Neon spectra Fig.~\ref{fig:Wavelegth_neon})})} (\text{FWHM}_{\text{ZEMAX}}) = 5.0~\text{pixels} \\
    \text{Spectral resolution} (\delta\lambda_{\text{ZEMAX}}) = \text{FWHM} \times \text{Dispersion}_{\text{ZEMAX}} = 5 \times 1.546 = 7.73~\text{\AA} \\
     \text{Resolving power} (R_{\text{ZEMAX},\lambda}) = \frac{\lambda}{\delta\lambda_{\text{ZEMAX}}} = \frac{6402.25}{7.73} \approx 828
\end{gather*}
\end{minipage}

\subsection{Laboratory Estimation of Spectral Resolution and Linear Dispersion}

After completing the ZEMAX simulations, the spectrograph was assembled and tested in the laboratory, as shown in Figure~\ref{fig:lab_spectrograph}. A total of 35 spectral traces were recorded on the detector, as two fibers were damaged during fabrication. The acquired data were initially in 2D format; the spectrum from a single fiber was extracted and converted into a 1D, wavelength calibrated spectrum using Neon emission lines (Figure~\ref{fig:Wavelegth_Space_lab_source}). Using this calibrated spectrum, the lab spectra results, including detector positions for wavelengths $\lambda$ and $\lambda_a$ and the resulting spectral performance metrics, are derived as follows:

\noindent
\begin{minipage}{\textwidth}
\abovedisplayskip=0pt
\belowdisplayskip=0pt
\begin{gather*}
    \text{Detector pixel positions for $\lambda$ and $\lambda_a$ is: } X_{\text{lab}} = 357.42~\text{pixels}, \quad X_{a,\text{lab}} = 343.57~\text{pixels} \\
    \text{Spatial separation between $\lambda$ and $\lambda_a$ on the detector } (\Delta X_{\text{lab}}) = \Delta X_{\text{lab}} = X_{\text{lab}} - X_{a,\text{lab}} = 13.85~\text{pixels} \\
    \text{Linear dispersion for laboratory data } (Dispersion_{\text{lab}}) = \frac{\lambda - \lambda_a}{\Delta X_{\text{lab}}} = \frac{19.26~\text{\AA}}{13.85~\text{pixels}} = 1.390~\text{\AA/pixel} \\
    \text{FWHM (determined from laboratory Neon spectra Fig.~\ref{fig:Wavelegth_neon}) } (\text{FWHM}_{\text{lab}})  = 5.6~\text{pixels} \\
    \text{Spectral resolution} (\delta\lambda_{\text{lab}}) = \text{FWHM}_{\text{lab}} \times Dispersion_{\text{lab}} = 5.6 \times 1.390 = 7.784~\text{\AA} \\
    \text{Resolving power} (R_{\text{lab},\lambda}) = \frac{\lambda}{\delta\lambda_{\text{lab}}} = \frac{6402.25}{7.784} \approx 822
\end{gather*}
\end{minipage}

\subsection{Comparison}

\begin{figure}[htbp]
  \centering
  \begin{subfigure}[b]{0.35\linewidth}
    \centering
    \includegraphics[width=\textwidth]{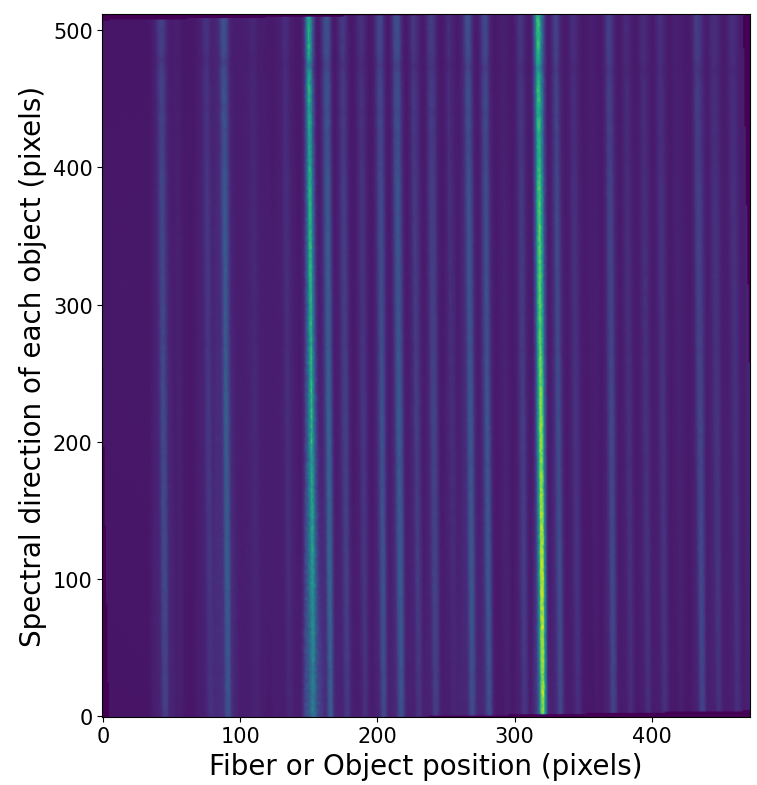}
    \caption{}
    \label{fig:2D_Spectra_Solar}
  \end{subfigure}
  \hspace{0.02\linewidth}
  \begin{subfigure}[b]{0.55\linewidth}
    \centering
    \includegraphics[width=\textwidth]{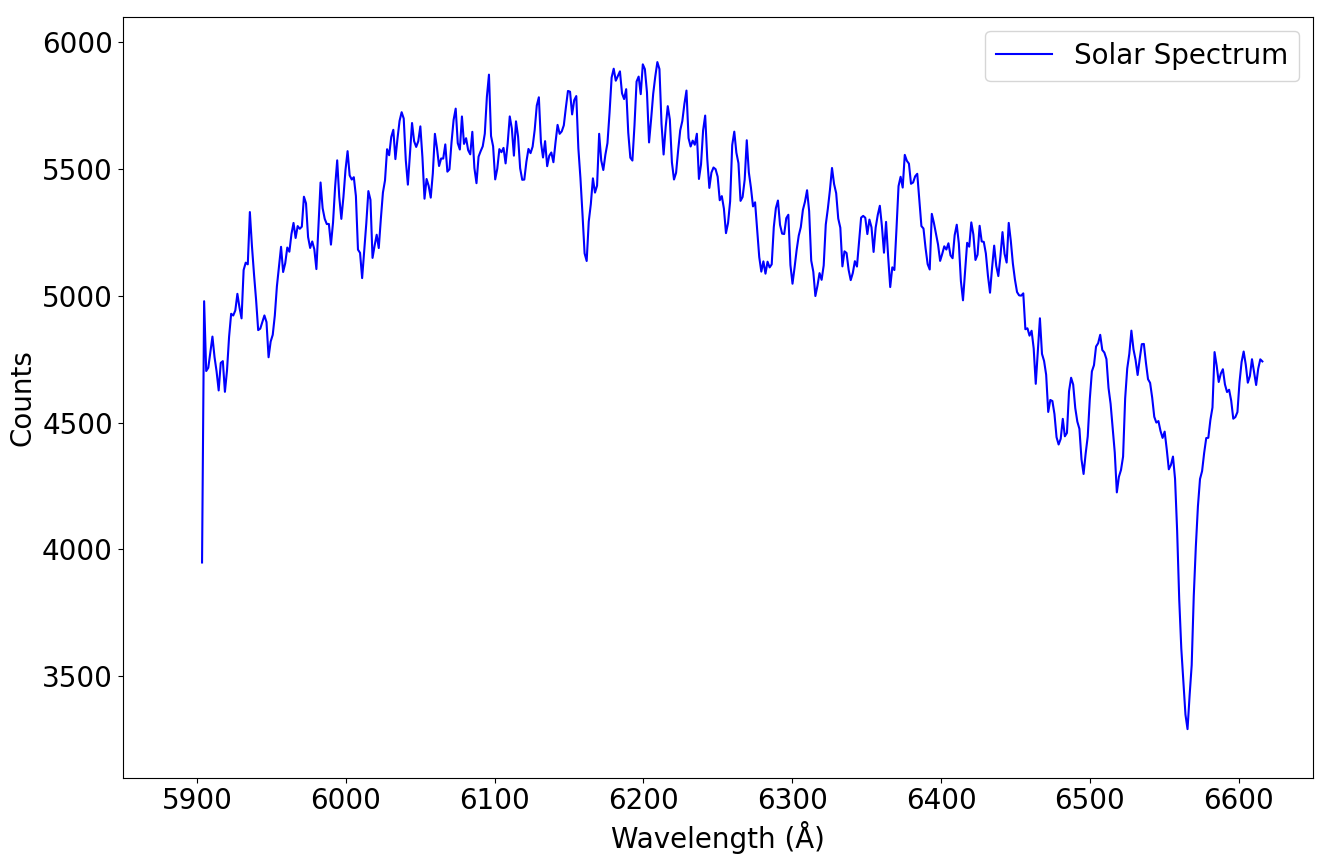}
    \caption{}
    \label{fig:Wavelegth_solar}
  \end{subfigure}

  \caption{Solar spectra at the IFU slit end. 
  (a) 2D spectrum of the solar source at the IFU slit end. 
  (b) Wavelength calibrated solar spectrum for a single object.}
  
  \label{fig:2D_Spectra_lab}
\end{figure}

The measured linear dispersion, spectral resolution, and resolving power show good agreement with the theoretical estimates, ZEMAX simulations, and laboratory IFS results (Table~\ref{tab:resolution_comparison}). Analysis of all laboratory fiber spectra yields a dispersion range of 1.38--1.42~\AA/pixel. The small deviation from the theoretical values is attributed to slight misalignment of the collimator and camera lenses, leading to a minor change in the effective system magnification. The IFU front end is then illuminated with solar light to further evaluate the system performance. The H-$\alpha$ absorption line is clearly identified at wavelength $\sim$6563~\AA (Figure~\ref{fig:2D_Spectra_lab}).

\begin{table}[htbp]
\centering
\caption{Comparison of linear dispersion, spectral resolution and resolving power for Theoretical, ZEMAX, and Laboratory Results}
\captionsetup{skip=0.1pt}
\label{tab:resolution_comparison}
\begin{tabular}{lccc}
\toprule
\textbf{Method} & \textbf{Linear Dispersion (\AA/pixel)} & \textbf{Spectral Resolution (\AA)} & \textbf{Resolving Power} \\
\midrule
Theoretical & 1.539 & 7.695 & 832 \\
ZEMAX Simulation & 1.546 & 7.73 & 828 \\
Laboratory & 1.39 & 7.784 & 822 \\
\bottomrule
\end{tabular}
\end{table}

\section{Conclusion}

We have successfully designed, fabricated, and laboratory tested a compact, fiber based IFS. The IFU was assembled using 37 fibers arranged in a hexagonal geometry to match the lenslet pupil pitch, achieving a placement accuracy of $<50~\mu\text{m}$, of which 35 fibers were effectively illuminated and characterized. The spectrograph was evaluated using both a Neon lamp and solar illumination. Linear dispersion, spectral resolution and resolving power were determined using theoretical calculations, ZEMAX simulations, and laboratory measurements, all of which are in close agreement. The H$\alpha$ line was clearly detected in the solar spectrum, with its measured wavelength in good agreement with the rest value. These results demonstrate good consistency between the optical design and experimental performance, establishing the feasibility of a compact, fiber based IFS for astronomical applications and providing a solid foundation for future implementation on the 2.34~m VBT.

\section*{Declarations}

\subsection{Availability of Data and Materials} \label{material_availibility}  

The data supporting the findings of this study are available from the corresponding author, Nitish Singh, upon reasonable request.   
  
\subsection{Author Contributions}

S. Sriram, Bharat Kumar Yerra, and Nitish Singh conceptualized the project. The optical design was carried out by Nitish Singh. Laboratory testing was performed by Nitish Singh and S. Sriram. The calculations and material selection were carried out by Nitish Singh, Jurgen Schmoll, and S. Sriram. Jurgen Schmoll provided guidance on arranging the fibers in a hexagonal geometry. The paper was written by Nitish Singh, Bharat Kumar Yerra, Jurgen Schmoll, and S. Sriram.

\subsection{Acknowledgements}

We sincerely thank Kathiravan S. and T. Vishnu for their assistance in arranging the needles in a hexagonal configuration. We also express our gratitude to Debadutta for assisting with the laboratory testing of the IFU. This research was supported financially by the Tata Consultancy Services Research Fellowship and the Indian Institute of Astrophysics, under the Department of Science and Technology, Government of India.

\subsection{Declaration of competing interest}

The authors declare no conflict of interest with any known people.

\bibliography{extra}
\bibliographystyle{bullsrsl-en}

\end{document}